\documentclass[nofootinbib, aps, prd, amsmath, floats, floatfix, twocolumn,superscriptaddress, showpacs,eqsecnum]{revtex4-1}

\usepackage{graphicx}
\usepackage{amsmath,amssymb}
\usepackage{amsfonts}
\usepackage{xspace} 
\usepackage[usenames]{color}
\usepackage{dcolumn}
\usepackage{bm}
\usepackage{hyperref}

\usepackage{ulem}
\newcommand{\eq}{\begin{equation}}
\newcommand{\eeq}{\end{equation}}
\newcommand{\be}{\begin{equation}}
\newcommand{\ee}{\end{equation}}
\newcommand{\bea}{\begin{eqnarray}}

\newcommand{\eea}{\end{eqnarray}}
\newcommand{\bes}{\begin{subequations}}
\newcommand{\ees}{\end{subequations}}

\begin{document}


\title{A Post-Newtonian approach to black hole-fluid systems}

\author{Enrico Barausse}
\affiliation{Institut d'Astrophysique de Paris, UMR 7095 du CNRS,
Universit{\'e} Pierre \& Marie Curie, 98bis Bvd. Arago, 75014 Paris, France}
\affiliation{Department of Physics, University of Guelph, Guelph, Ontario, N1G 2W1, Canada}

\author{Luis Lehner}
\affiliation{Perimeter Institute for Theoretical Physics, Waterloo, Ontario N2L 2Y5, Canada}


\begin{abstract} 
This work devises a formalism to obtain the equations of motion for a black hole-fluid configuration. 
Our approach is based on a Post-Newtonian expansion and adapted  to
scenarios where obtaining the relevant dynamics requires long time-scale
evolutions. These systems are typically studied with Newtonian approaches, which have the advantage that
larger time-steps can be employed than in full general-relativistic  simulations, but
have the downside that important physical effects are not accounted for. The formalism presented
here provides a relatively straightforward way to incorporate those effects in existing implementations,
up to 2.5PN order, with lower computational costs than fully relativistic  simulations.
\end{abstract}

\date{\today \hspace{0.2truecm}}

\pacs{04.25.dg, 04.25.Nx, 97.60.Lf}

\maketitle

\section{Introduction}
The interaction between black holes and matter in accreting systems is
known to power emission across a wide spectrum, and provides the engine for 
exciting phenomena such as AGNs, blazars, quasars, etc. Systems where a black hole interacts
with a localized matter distribution also lie at the heart of other interesting
astrophysical processes. For instance, black hole-star systems can give rise
to strong tidal interactions that might induce supernova-like events~\cite{Rosswog:2008hq};  the disruption of stars by a 
supermassive
black hole can trigger strong  
flares~\cite{MacLeod:2012cr,Bloom:2011xk,Kasen:2009cv,Hayasaki:2012ia,Stone:2010sr,Zalamea:2010mv};
stellar-mass black holes interacting with a neutron star can be responsible for short 
gamma-ray bursts~\cite{Piran:2012wd,Metzger:2010sy}; etc. 
The upcoming close encounter of our own SgrA$^*$ with a gas cloud (G2), expected for next 
year~\cite{Gillessen:2011aa}, will provide unprecedented opportunities to study a nearby example of 
such interactions.
 
The understanding of these systems requires dealing with diverse physics -- gravity and matter at least --  
and therefore solving general relativistic hydrodynamic equations in dynamical, strongly gravitating scenarios.
Due to the complex and non-linear character of the underlying equations, numerical simulations are required to make realistic 
headways into the problem.
This task, however, is formally complicated by several issues. First, depending on the problem
under consideration,  vastly different scales are involved -- star and black-hole sizes --,
as well as, potentially, extremely long dynamical times. Second, the physical theories 
one has to employ -- general relativity and relativistic hydrodynamics -- are not only highly involved/non-linear, but also bring different
characteristic timescales for propagating modes. In particular,  general relativity dictates that gravitational
degrees of freedom propagate at the speed of light, while hydrodynamic modes do so at the sound speed, which
can be significantly lower. This implies, at a computational level, that a fully general-relativistic study
is necessarily more costly than a non-relativistic one.

There exist, however, a class of problems where propagating gravitational effects are secondary, and the above-mentioned limitation 
can be dealt with by suitable approximations. A typical approach is to assume that gravitational
effects are described by Newtonian gravity, where the gravitational 
field is determined via elliptic equations, which can be solved efficiently (e.g.~\cite{multigrid1,multigrid2,multigrid3}) 
once appropriate boundary conditions are specified.
The only hyperbolic equations are then given by hydrodynamics and, as a consequence, the dynamical propagation 
speeds are given by the speed of sound, which allows considerably larger step-sizes to be adopted.
Gravitational radiation reaction may then be approximately included by using the quadrupole formula or higher-order extensions of
it (c.f. for instance Ref.~\cite{r_instability,2009ApJ...705..844G,2010AA...514A..66R,2006Sci...312..719P}).

However, for scenarios involving a black hole -- even when ignoring propagating modes --, this strategy must
be modified to incorporate crucial features brought by general relativity, which cannot be ignored unless the star
 remains very far from the black hole through the whole regime of interest.
A widely used approach to address this issue is to consider the gravitational potential 
sourced by the star, and add to it the so-called
``Pacynski-Wiita'' (PV)  potential $V(r)=-GM/(r-2 GM/c^2)$~\cite{1980AA....88...23P}. This method therefore amounts to a phenomenological modification to
the Newtonian potential, and guarantees the presence of an innermost stable circular orbit (ISCO)~\cite{2009AA...500..213A}.
The PV potential, however, does not depend on the black-hole spin, and thus cannot properly describe rotating black holes.
Furthermore, even for non-spinning black holes, the ISCO has different orbital frequency and angular momentum than in a Schwarzschild 
spacetime, even though it lies at $r=6 M$, as for a Schwarzchild spacetime in the usual areal coordinates.
Various increasingly sophisticated suggestions  have been presented to partially address these shortcomings
(see e.g.~\cite{1991ApJ...378..656N,1999AA...343..325S,2002MNRAS.335L..29K,Wegg:2012yx,Tejeda:2013mva}) for
non-spinning black holes. However, crucial features brought by the presence of a black hole are still not fully accounted 
for (e.g. gravitational redshift, frame-dragging, gravitational radiation, etc.) unless further ad-hoc ingredients are incorporated.
Naturally, since strong observational evidence indicates that gravitational waves exist and carry energy and angular
momentum away from the system, as well as that black holes possess a variety of spins (spanning the whole allowed range 
$cJ/(GM^2) \leq1$,  see e.g. Ref.~\cite{McClintock:2011zq}), a better treatment of black hole effects 
on matter is desirable.

In this work, motivated by the above observations, we describe a systematic approach to treat the problem starting from
the correct description in full General Relativity, and introduce a post-Newtonian (PN) expansion to capture relativistic effects
in a consistent manner. This goal, as we describe below, has been pursued by other authors -- either via fully
general relativistic or approximate methods -- but the resulting approaches require significantly revamping/altering
traditional astrophysics modeling strategies. We thus adopt an approach that is intimately tied with
the traditional ``Newtonian-route'', providing a systematic way to include further physical ingredients in a controlled expansion
of the equations or, alternatively, to estimate errors in the existing approaches. 

For context, we briefly review related work.
For instance, Ref.~\cite{BDS_hydro} developed a formalism similar to ours, focusing on fluid systems only (i.e. without
a black hole) in the Arnowitt-Deser-Misner (ADM) gauge, with leading-order dissipative effects, and conservative effects
included at 1PN order (see also Ref.~\cite{faye_higher_order_fluxes} for an extension of 
this formalism to next-to-leading order in the dissipative sector). The numerical implementation of the formalism of Ref.~\cite{BDS_hydro}, 
performed in Refs.~\cite{ayal,Oohara_nakamura1,Oohara_nakamura2},  
achieved promising results, but found difficulties to produce sufficiently massive stars to describe
neutron stars~\cite{rosswog_private}. As we illustrate in this work, to deal with these difficulties higher-order approximations are required. 
An attempt in this direction was made by Ref.~\cite{shibata}, which performed a comprehensive study of the equations of PN hydrodynamics and 
radiation reaction (for fluid systems only, i.e. in the absence
of a black hole) in the 3+1 formalism and with several gauge choices, through 2.5PN order.

Higher-order extensions are particularly easy to achieve in the harmonic gauge often used in PN calculations,
but this gauge choice would yield a scheme
involving hyperbolic equations with characteristics speeds given by the speed of light, already at low PN orders.\footnote{In the harmonic gauge,
all of the metric perturbations satisfy wave equations, although the effects of gravitational-wave emissions on 
 observable quantities only appear at 2.5PN order, as expected.}
 Such behavior is undesirable
for studying mildly dynamical spacetimes whose 
true dynamics is governed
by the characteristic speeds of the fluid describing the matter content.
Along these lines, a recently introduced formalism, partially related to ours, is that of Refs.~\cite{kim1,kim2}, 
which essentially consists of solving the 1PN Einstein equations with a conformally flat ansatz for the metric. 
While this approach indeed gets rid of characteristics with light-speed propagation, it only accounts for \textit{some}
effects at 1PN.

A different approach to describe compact-object binaries is given by the effective-one-body formalism (EOB)~\cite{BD99},
which attempts to accelerate the convergence of the PN dynamics by ``resumming'' it (both
in the conservative~\cite{BD99,DJS3PN,BBL12,damour01,DJS,BB10,BB11,nagar} and dissipative~\cite{DamourResummedWfms,tagoshi} 
sectors). While very successful
at describing binary black holes (see e.g. Ref.~\cite{taracchini} for a recent comparison to full general-relativistic results) and more recently
neutron-star binaries~\cite{EOB_NS1,EOB_NS2}, the EOB is not suitable for describing black holes interacting with
matter that is not in a  compact configuration. Last, within the fully general relativistic regime a few options have
been pursued to study related systems within reasonable computational efforts. For instance, in~\cite{East:2013iwa} a method
to reduce the computational cost of implementing the full Einstein equations coupled to matter has recently
been introduced. This method allows one to treat black hole-star binary systems where the backreaction on the black hole
is sufficiently small. In another approach, to study stellar black hole-white dwarf systems a suitable rescaling of the equation of state
was introduced to study a related, though computationally tractable, problem and then extrapolate results to the problem of
interest~\cite{Paschalidis:2010dh}. Alternatively, focus has been placed in the interaction of white dwards with
intermediate mass black holes  so as to deal with comparable scales~\cite{Shcherbakov:2012iv}. However, even in this regime simulations
could only track the system for relatively short times.

While keeping in mind these issues and options, we here develop an approach consisting of minimal modifications to 
Newtonian theory. This has the advantage that existing (thoroughly tested/highly sophisticated) 
implementations of hydrodynamics (for representative examples, see e.g.~\cite{2010AA...514A..66R,2009ApJ...705..844G,2013ApJ...767...25G})
in Newtonian theory could be easily modified to implement our approach. Alternatively, 
our approach can be used to evaluate the importance of the
PN effects that are missing in Newtonian simulations, and therefore gauge their errors relative to a fully general-relativistic treatment.

This work is organized as follows. Section \ref{secbasic} describes our basic strategy for adapting
the standard PN expansion to the purpose of describing a black hole interacting
with a relativistic-fluid configuration. In sections \ref{sec1PN}  and \ref{sec2PN} we present the derivations
of the equations to first and second PN order. Section \ref{sec2.5PN} describes how gravitational waves are accounted
for, and how their effect can be incorporated in the elliptic system determining the gravitational and fluid behavior.
In section \ref{secKERRFLUID} we describe how to account for the black hole presence, while in section \ref{secMODELS} we
present a discussion of
implementation choices, together with a few examples. 

\section{Basic strategy}\label{secbasic}

An analytic description of the two-body dynamics in General Relativity, 
dating back originally to Einstein's calculation of the perihelion of Mercury~\cite{mercury},
 is given by the PN approximation.
The PN formalism is essentially an expansion in the ratio between the 
typical velocity of the system, $v$, and the speed of light (see Ref.~\cite{blanchet} for a recent review).
For a fluid system, the dynamics is known 
through order $(v/c)^5$ beyond Newtonian theory (i.e. through 2.5PN order)~\cite{PNfluid1,PNfluid2,PNfluid3}.
For a binary system of non-spinning black holes, the conservative dynamics is known 
through 3PN order~\cite{Da.al.01,Bl.al.04}, while the gravitational-wave fluxes
have been computed through 3.5PN order \cite{Bl.al2.05,Ki.08,Bl.al.08}  (i.e. through order $(v/c)^7$
beyond the quadrupole formula) and 3PN order \cite{Ar.al2.09}, respectively for a system of two non-spinning black holes on circular 
or eccentric orbits. The effect of the spins on the dynamics of compact-object (and in particular black-hole) binaries has also been calculated, both in the conservative
and dissipative sectors~\cite{Bl.al.06,Bl.al2.11,Da.al.08,St.al.08,St.al2.08,HeSc2.08,HaSt.11,Ki.95,Fa.al.06,FoSt.11,Po.al2.11,PoRo.06,
  Po.06,PoRo.08,PoRo2.08,Po.10,Le.10,Le2.11}. 

One drawback of the PN expansion is that it is slowly convergent. As a result, the PN dynamics, if extrapolated
to $v\sim c$, does not reproduce the correct general-relativistic
dynamics accurately. This is particularly true for binary black-hole systems near coalescence. In fact, in
order to obtain a sensible description of such systems in the strong field regime, the PN equations need to be ``resummed''
(i.e. completed by ``educated guesses'' of the higher-order terms in $v/c$, based on the known lower-order ones), both
in the conservative~\cite{BD99,DJS3PN,BBL12,damour01,DJS,BB10,BB11,nagar} and dissipative~\cite{DamourResummedWfms,tagoshi} sectors. 
This resummation results in the so-called EOB model, which was originally proposed in Ref.~\cite{BD99}, and which works not only for binary black holes~\cite{taracchini}, 
but can also be adapted to more general compact-object binaries~\cite{EOB_NS1,EOB_NS2}. 

Seeking a reasonable compromise between the Newtonian (or pseudo-Netwonian) calculations commonly performed in astrophysics,
and a fully general relativistic approach used in gravitational-wave physics, 
we therefore propose a formalism based on the PN dynamics through 2.5PN order (i.e. 2PN order in the conservative sector, 
leading order in the dissipative sector).
This generalizes the 1PN conservative, leading-order dissipative model of Ref.~\cite{BDS_hydro}, except that \text{(i)} our model  allows
for the presence of
a spinning black hole, besides the fluid; \text{(ii)} while not fully resummed
into an EOB model, we ``resum'' at the least the energy and rest-mass conservation equations, as well as the Euler equation, and show that this 
resummation significantly improves the behavior of our model.

At a practical level, and in order to obtain a model causally allowing for larger
timesteps than those implied by a fully general-relativistic implementation, 
we seek a formalism in which the PN Einstein equations neatly separate into
a system of elliptic equations for ``gravitational potentials'' and hyperbolic evolution equations for the
matter variables. 
(A formalism allowing for a similar decomposition has been obtained, at least partially, also in the fully general-relativistic case, 
where it is known as ``fully-constrained formulation''  of the Einstein equations~\cite{fully_constrained}). 
With such a decomposition, 
the elliptic equations will provide the generalization of the Poisson equation
for the Newtonian potential, the equation for the frame-dragging potential, etc. 

To achieve this separation, we draw inspiration from perturbation theory. Consider a
metric perturbation $h_{\mu\nu}$ over a generic curved background metric $g_{\mu\nu}$,
and introduce the trace-reversed perturbation
\begin{equation}
\label{rev_trace}
\bar{h}_{\mu\nu}\equiv
h_{\mu\nu}-\frac12\,h^\alpha_\alpha\,g_{\mu\nu}\,.
\end{equation}
In the Lorenz gauge 
$\nabla_\nu \bar{h}^{\mu\nu}=0$, the linearized Einstein equations
take the  deceivingly simple form~\cite{Poisson_lorenz,Sciama_lorenz}
\begin{equation}\label{eq:box_h}
\Box\, \bar{h}^{\alpha\beta} + 2 R_{\mu\ \nu}^{\ \alpha\ \beta}
	\bar{h}^{\mu\nu} + S_{\mu\ \nu}^{\ \alpha\ \beta}
        \bar{h}^{\mu\nu} = 
	-16\pi T_{\rm fluid}^{\alpha\beta}\;,
\end{equation}
where
\begin{gather}
S_{\mu\alpha\nu\beta} = 2 G_{\mu(\alpha} g_{\beta)\nu} -
R_{\mu\nu} g_{\alpha\beta}  - 2 g_{\mu\nu} G_{\alpha\beta}\;,\\
\Box\equiv g^{\mu\nu}\nabla_\mu\nabla_\nu
\end{gather}
($R_{\mu\nu\alpha\beta}$, $R_{\mu\nu}$ and $G_{\mu\nu}$ being the
background Riemann, Ricci and Einstein tensors, and $\nabla$ being the background Levi-Civita connection). The fluid's stress-energy 
tensor satisfies the linearized conservation equation,
\begin{multline} -16\pi\, { \nabla_\beta\,
    T^{\alpha\beta}_{\rm
      fluid}}=2G^{\beta\sigma}\nabla_\sigma\bar{h}_{\beta}^\alpha\\-2
  G^{\alpha\beta}\partial_\beta\bar{h}-
	R^{\beta\sigma}\nabla_\gamma\bar{h}_{\beta\sigma}g^{\gamma\alpha}\;.
\label{eq:euler}
\end{multline}
(see e.g. Ref.~\cite{BHtorus}).
In spite of its apparent simplicity, eq. (\ref{eq:box_h}) has
all of the metric perturbations propagating at the speed of light, and would therefore share the same 
time-stepping constraint of the full general-relativistic treatment. 

A gauge that is better suited to our purposes, which include providing a formalism
based on elliptic equations,
is the so-called Poisson gauge used in 
cosmology~\cite{bertschinger,bombelli,scott_lect}.
(Note that this gauge can be chosen also at non-linear orders, see e.g. Ref.~\cite{bruni}). Focusing on
perturbations over a flat background, we can write the most generic perturbed flat metric in 
Cartesian coordinates $x^0=ct$, $x^i$ ($i=1,2,3$) as\footnote{Note that eq.~\eqref{metric} is a \textit{definition}. For instance, one {\it defines}
$g_{00}$ to be $-\left( 1+2{\phi}/{c^2}\right)$. Obviously, higher-order (quadratic) terms in $\phi$, $\psi$, etc will then show up in the field equations, because
of the non-linear character of the Einstein equations.}
\begin{align}\label{metric}
&g_{00}=-\left( 1+2 \frac{\phi}{c^2}\right)\, ,\nonumber\\
&g_{0i}= \frac{\hat{\omega}_i}{c^3}
\, ,  \nonumber\\&g_{ij}=\left(1 -2 \frac{\psi}{c^2}\right)\delta_{ij}+
 \frac{\hat{\chi}_{ij}}{c^2} 
\,,
\end{align}
where the perturbations can be decomposed into scalar parts, transverse 
({\it i.e} divergence-free) vector parts, and transverse trace-free (TT) tensor 
parts as
\begin{equation}
\hat{\omega}_i=\partial_i\omega+\omega_i\, ,
\end{equation}
\begin{equation}
\hat{\chi}_{ij}=\left(\partial_{ij}-\frac13 \delta_{ij} \nabla^2\right)\chi+\partial_{(i}\chi_{j)}
+\chi_{ij}\,,
\end{equation}
where $\partial_i{\omega}^i=\partial_i{\chi}^{i}=\partial_i{\chi}^{ij}={\chi}^{i}_i=0$.
(The indices are raised and lowered with the flat metric). The Poisson gauge is then defined by the conditions 
$\partial_i\hat{\omega}^i=\partial_i\hat{\chi}^{ij}=0$, which imply $\omega=\chi=\chi_{i}=0$. 
The metric is then given by
\begin{align}\label{metric2}
&g_{00}=-\left( 1+2 \frac{\phi}{c^2}\right)\, ,\nonumber\\
&g_{0i}= \frac{{\omega}_i}{c^3}
\, ,  \nonumber\\&g_{ij}=\left(1 -2 \frac{\psi}{c^2}\right)\delta_{ij}+
 \frac{\chi_{ij}}{c^2}
\,,
\end{align}
with  $\partial_i{\omega}^i=\partial_i{\chi}^{ij}={\chi}^{i}_i=0$, and the perturbed Einstein
equations, at the linear order, will reduce to elliptic equations\footnote{This can be checked
by confirming that the symbol of the principal part of the system has non-vanishing determinant for non-zero
vectors, see Ref.~\cite{Dain:2004nt}.} for the potentials
$\phi$, $\psi$ and $\omega_i$,~\cite{bertschinger,bombelli,scott_lect,bardeen_pert}
\begin{gather}
\nabla^2\phi= \ldots\label{cartoon1}\\
\nabla^2\phi= \ldots\label{cartoon2}\\
\nabla^2\omega_i= \ldots\label{cartoon3}
\end{gather}
(the dots indicating the source terms),
while the TT perturbation $\chi_{ij}$ (which has only two degrees of freedom, representing the
two polarizations of the graviton) will satisfy a wave equation~\cite{bertschinger,bombelli,scott_lect,bardeen_pert}
\begin{equation}
\Box \chi_{ij}=\ldots\label{cartoon4}
\end{equation}

Therefore, when deriving the PN equations, we will choose to \textit{not} use the standard harmonic gauge used in PN theory,
because that is very similar to the Lorenz gauge introduced above, i.e. it leads to hyperbolic equations for all of the metric perturbations. 
Instead, we adopt the Poisson gauge, and show that it leads to PN equations that have a 
structure similar to eqs.
\eqref{cartoon1}--\eqref{cartoon4}. While the source terms of the PN equations
will be rather involved (although straightforward to implement),
the elliptic equations  \eqref{cartoon1}-\eqref{cartoon3} can be solved at arbitrary times and so the overall
time-stepping criterion is determined by the speed of sound.
As for the solution to the wave equation \eqref{cartoon4}, we will show that at leading order it 
is given by the solution of another elliptic equation, with the addition of another contribution involving time derivatives 
of numerical integrals.

\section{The equations at 1 PN order}\label{sec1PN}

As mentioned in the previous section, we start from the perturbed Minkowski metric in
the Poisson gauge, given by eq.~\eqref{metric2}. Also, we assume that the perturbations are sourced by a perfect fluid with mass-energy density $\rho$,
pressure $p$, and 4-velocity $u^\mu$ such that
\begin{equation}
\frac{u^i}{u^0}=\frac{v^i}{c}\,.
\end{equation}
[This implies $u^0=1-(\phi-v^2/2)/c^2+{\cal O}(1/c^4)$ because of the 4-velocity normalization.]

The Einstein field equations are, as usual,
\begin{equation}
G_{\mu\nu}=\frac{8\pi}{c^4} T_{\mu\nu}\,,
\end{equation}
or equivalently 
\begin{equation}
R_{\mu\nu}=\frac{8\pi}{c^4} \left(T_{\mu\nu}-\frac12 T g_{\mu\nu}\right)\,,
\end{equation}
where 
\begin{equation}
T^{\mu\nu}= (p+\rho c^2) u^\mu u^\nu +p g^{\mu\nu}
\end{equation}
is the fluid's stress-energy tensor, and where we are setting $G=1$ (as in the rest of this paper).
It is convenient to project the field equations onto tetrads carried by the fluid elements. 
More specifically, introducing the projector
$h_{\mu\nu}=g_{\mu\nu}+u_\mu u_\nu$, we consider the following equations:
\begin{gather}
E_{(0)(0)} \equiv \left(G_{\mu\nu}-\frac{8\pi}{c^4} T_{\mu\nu}\right) u^\mu u^\nu =0\\
E_{(0)(i)} \equiv \left(G_{\mu\nu}-\frac{8\pi}{c^4} T_{\mu\nu}\right) u^\mu h_i^\nu =0\\
E_{(i)(j)} \equiv \left[R_{\mu\nu}-\frac{8\pi}{c^4} \left(T_{\mu\nu}-\frac12 T g_{\mu\nu}\right)\right] h^\mu_i h_j^\nu =0\;.\label{Eij}
\end{gather}
Likewise, the stress-energy tensor
conservation $\nabla_{\mu} T^{\mu\nu}=0$ implies the energy conservation equation
\begin{equation}
u^\mu \partial_\mu \rho=-\left(\frac{p}{c^2}+\rho\right) \theta\,,\qquad \theta\equiv \nabla_\mu u^\mu
\end{equation}
when projected along the 4-velocity $\boldsymbol u$, and the Euler equation
\begin{equation}
u^\nu \nabla_\nu u^\mu =-\frac{h^{\mu\nu}\partial_{\nu} p}{p+\rho c^2}
\end{equation}
when projected on the hyperplane orthogonal to $\boldsymbol u$. Finally,
the rest-mass conservation equation is
\begin{equation}
\nabla_{\alpha} (\mu u^\alpha)=0\,,
\end{equation}
where $\mu$ is the rest-mass density.

The TT part of $E_{(i)(j)}=0$ immediately gives 
\begin{equation}
\chi_{ij}={\cal O}\left(\frac{1}{c^2}\right)\,,
\end{equation}
which is hardly surprising because it simply 
amounts to the absence of gravitational waves at this order of approximation.
The 
off-diagonal part of $E_{(i)(j)}=0$ then gives 
\begin{equation}\label{phi_eq_psi}
\psi=\phi+{\cal O}\left(\frac{1}{c^2}\right)\,,
\end{equation}
and we can then define 
\begin{equation}
\psi\equiv\phi+\frac{\delta\psi}{c^2} \,.
\end{equation}
From $E_{(0)(i)}=0$, using $E_{(0)(0)}=0$ at the lowest order in $1/c$ and eq.~\eqref{phi_eq_psi}, we get an
equation for the ``frame-dragging'' potential:
\begin{equation}\label{nabla2wi1PN}
\nabla^2 \omega^i= 4 (4 \pi\rho v^i+\phi_{,ti})+{\cal O}\left(\frac{1}{c^2}\right)\,,
\end{equation}
while taking a linear combination of  $E_{(0)(0)}=0$
and $\delta^{ij}E_{(i)(j)}=0$ to eliminate $ \delta\psi$, we obtain
\begin{multline}
\nabla^2 \phi=4\pi \left(3 \frac{p}{c^2} +\rho\right)+\frac{2}{c^2}{\phi}_{,i}{\phi}_{,i}\\
+8 \pi\rho \left(\frac{v}{c}\right)^2-\frac{3}{c^2}{\phi}_{,tt}+{\cal O}\left(\frac{1}{c^4}\right)\,.
\end{multline}

The relativistic energy-conservation and Euler equations give
\begin{multline}\label{rhodot1PN}
{\rho}_{,t}+{\rho}_{,i}v^{i}+\left(\frac{p}{c^2}+\rho\right){v^{i}}_{,i}-\frac{p_{,i} v^i}{c^2} 
\\- \frac{\rho}{c^2} (4 \phi_{,i} v^i+3 \phi_{,t})={\cal O}\left(\frac{1}{c^4}\right)\,,
\end{multline}
and
\begin{multline}\label{vidot1PN}
{v^{i}}_{,t}+{v^{i}}_{,a}v^{a} =
-{\phi}_{,i}
-\frac{{p}_{,i}}{\rho+p/c^2}  
-\frac{4 \phi \,{p}_{,i}}{\rho c^2}
-\frac{2 \phi \,\phi_{,i}}{c^2} \\
 - \left(\frac{v}{c}\right)^2 \left(-\frac{p_{,i}}{\rho}+\phi_{,i}\right)
+\frac{v^i}{c^2}  \left(4\phi_{,j} v^j+3\phi_{,t}-\frac{p_{,t}}{\rho}\right)\\+
\frac{1}{c^2}\left(
-{\omega^{i}}_{,t}
-{\omega^{i}}_{,a}v^{a}
+{{\omega^{a}}}_{,i}v^{a}\right)+{\cal O}\left(\frac{1}{c^4}\right)
\,,
\end{multline}
while the rest-mass conservation equation, combined with the energy-conservation equation, gives
\begin{equation}\label{Udot1PN}
\partial_t U=\ -\left(p+U\right){v^{j}_{ }}_{,j}-{U}_{,j}v^{j}+{\cal O}\left(\frac{1}{c^2}\right)\,,
\end{equation}
where $U$ is the internal-energy density, defined by $\rho=\mu+U/c^2$.

Finally, we note that because of the energy conservation at Newtonian order, the right-hand side of eq.~\eqref{nabla2wi1PN} has zero divergence. Therefore,
$\nabla^2(\partial_i\omega^i)={\cal O}\left({1}/{c^2}\right)$, and
the gauge condition $\partial_i \omega^i=0$ is automatically satisfied, at 1PN order, if $\partial_i \omega^i$ goes to zero far from the source. 

\section{The equations at 2 PN order}\label{sec2PN}

Going to the next order in $1/c^2$, it is straightforward (although laborious) to obtain the 2PN equations. More specifically,
replacing the 1PN equations derived in the previous section into $E_{(0)(0)}$ and $\delta^{ij}E_{(i)(j)}=0$, we obtain
\begin{eqnarray}
&&\nabla^2 \phi=4\pi \left(3 \frac{p}{c^2} +\rho\right)+\frac{2}{c^2}{\phi}_{,i}{\phi}_{,i}
+8 \pi\rho \left(\frac{v}{c}\right)^2-\frac{3}{c^2}{\phi}_{,tt}\nonumber\\
&&
+\frac{1}{c^4}\Big[-16\pi\rho{}{\phi{}}^{2}-8\pi\rho{}\delta\psi{}+{\phi{}}_{,i}{\delta\psi{}}_{,i}+{\phi{}}_{,i}{\omega{}^{i}_{ }}_{,t}
-\frac12 {\omega{}^{i}_{ }}_{,j}{\omega{}^{i}_{ }}_{,j}\nonumber\\
&&
+\frac12 {\omega{}^{i}_{ }}_{,j}{\omega{}^{j}_{\ }}_{,i}+8\pi(p-4\rho{}\phi{})v^2+8\pi\rho v^4
+{\phi{}}_{,ij}\chi{}^{ij}_{  }\nonumber\\
&&-3{\delta\psi{}}_{,tt}\Big]+{\cal O}\left(\frac{1}{c^6}\right)\,,
\end{eqnarray}
\begin{multline}
\nabla^2\delta\psi=-12p\pi-16\pi\rho{}\phi{}-{7\over 2}{\phi{}}_{,j}{\phi{}}_{,j}\\-4\pi\rho{}v^2
+3{\phi{}}_{,tt}+{\cal O}\left(\frac{1}{c^2}\right)\,.
\end{multline}
From the traceless part of $E_{(i)(j)}=0$, we obtain
\begin{eqnarray}
&&
\nabla^2\chi_{ij}=\frac{1}{c^2}\Big[\delta^{ij}(8\pi p+{\phi{}}_{,k}{\phi{}}_{,k}+8\pi\rho{}v^2-2{\phi{}}_{,tt})\nonumber\\
&&
+4{\phi{}}_{,i}{\phi{}}_{,j}+8\phi{}{\phi{}}_{,ij}+2{\delta\psi{}}_{,ij}
-{\omega{}^{i}_{ }}_{,jt}-{\omega{}^{j}_{ }}_{,it}-16\pi\rho{}v^{i}_{\  }v^{j}_{ }\Big]\nonumber\\
&&+{\cal O}\left(\frac{1}{c^4}\right)\label{eq:nabla2chi_ij}\,,
\end{eqnarray}
while from $E_{(0)(j)}=0$ it follows that
\begin{eqnarray}
&&\nabla^2 \omega^i= 4 (4 \pi\rho v^i+\phi_{,ti})
+\frac{2}{c^2}\Big[{\phi{}}_{,j}{\omega{}^{j}_{ }}_{,i}-{\phi{}}_{,ij}\omega{}^{j}_{\  }
\nonumber\\
&&
+2\left({\phi{}}_{,t}{\phi{}}_{,i}+{\delta\psi{}}_{,it}+4p\pi v^{i}_{\  }-16\pi\rho{}\phi{}v^{i}_{ }
+4\pi\rho{}v^2 v^{i}_{ }+2\pi\rho{}\omega{}^{i}_{ }\right)\Big]
\nonumber
 \\
 &&+{\cal O}\left(\frac{1}{c^4}\right)\,.
\end{eqnarray}
The energy conservation equation yields
\begin{eqnarray}\label{rhodot2PN}
&&{\rho}_{,t}+{\rho}_{,i}v^{i}+\left(\frac{p}{c^2}+\rho\right){v^{i}}_{,i}-\frac{p_{,i} v^i}{c^2} 
- \frac{\rho}{c^2} (4 \phi_{,i} v^i+3 \phi_{,t})=\nonumber\\&&
\frac{1}{c^4}\Big[
4(p+\rho{}\phi{}){\phi{}}_{,i}v^{i}_{ }+3\rho{}{\delta\psi{}}_{,i}v^{i}_{\ }-\rho{}{\omega{}^{i}_{ }}_{,j}v^{i}_{ }v^{j}_{ }+{p}_{,i}\omega{}^{i}_{ }
+\rho{}{\phi{}}_{,i}\omega{}^{i}_{\  }
\nonumber\\
&&+3p{\phi{}}_{,t}+6\rho{}\phi{}{\phi{}}_{,t}+v^2({p}_{,t}-\rho{}{\phi{}}_{,t})
+3\rho{}{\delta\psi{}}_{,t}\Big]+{\cal O}\left(\frac{1}{c^6}\right)\,,\nonumber\\&&
\end{eqnarray}
and the Euler equation becomes
\begin{eqnarray}\label{vidot2PN}
&&{v^{i}}_{,t}+{v^{i}}_{,a}v^{a} =
-{\phi}_{,i}
-\frac{{p}_{,i}}{\rho+p/c^2}  
-\frac{4 \phi \,{p}_{,i}}{\rho c^2+p}\nonumber\\
&&
-\frac{2 \phi \,\phi_{,i}}{c^2} 
 - \left(\frac{v}{c}\right)^2 \left(-\frac{p_{,i}}{\rho+p/c^2}+\phi_{,i}\right)\nonumber\\
&&
+\frac{v^i}{c^2}  \left(4\phi_{,j} v^j+3\phi_{,t}-\frac{p_{,t}}{\rho+p/c^2}\right)\nonumber\\
&&+
\frac{1}{c^2}\left(
-{\omega^{i}}_{,t}
-{\omega^{i}}_{,a}v^{a}
+{{\omega^{a}}}_{,i}v^{a}\right)\nonumber\\
&&+\frac{1}{c^4}
\Big[
-2\phi{}{\omega{}^{i}_{ }}_{,j}v^{j}_{ }+2\phi{}{\omega{}^{j}_{\ }}_{,i}v^{j}_{ }-{\chi{}^{ij}_{  }}_{,t}v^{j}_{ }
-{\chi{}^{ij}_{ \  }}_{,k}v^{j}_{ }v^{k}_{ }
\nonumber\\
&&
+{1\over 2}{\chi{}^{jk}_{  }}_{,i}v^{j}_{ }v^{k}_{\  }+\frac{{p}_{,j}}{\rho+p/c^2}(\chi^{ij}+\omega{}^{j}_{ }v^{i}_{\  })+{\phi{}}_{,j}\chi{}^{ij}_{ \  }
\nonumber\\
&&
-2(\delta\psi{}+4 {\phi{}}^{2}-v^{j}_{ }\omega{}^{j}_{ })\frac{{p}_{,i}}{\rho+p/c^2}
-4{\phi{}}^{2}{\phi{}}_{,i}-2\delta\psi{}{\phi{}}_{,i}\nonumber\\
&&
-2\phi{}{\omega{}^{i}_{\  }}_{,t}
+2{\delta\psi{}}_{,j}v^{j}_{ }v^{i}_{ }
-{\omega{}^{j}_{ }}_{,k}v^{j}_{ }v^{k}_{ }v^{i}_{\  }
+{\phi{}}_{,j}\omega{}^{j}_{ }v^{i}_{ }
+2\phi{}{\phi{}}_{,t}v^{i}_{ }
\nonumber\\
&&
+2{\delta\psi{}}_{,t}v^{i}_{ }
+v^2\Big(-2\phi{}{\phi{}}_{,i}-{\delta\psi{}}_{,i}
+\frac{{\  p}_{,t}}{\rho+p/c^2}v^{i}_{ }
-{\phi{}}_{,t}v^{i}_{ }\Big)\nonumber\\
&&
+2{\phi{}}_{,j}v^{j}_{ }\omega{}^{i}_{ }
-\frac{{p}_{,t}}{\rho+p/c^2}\omega{}^{i}_{ }+{\phi{}}_{,t}\omega{}^{i}_{ }\Big]+{\cal O}\left(\frac{1}{c^6}\right)\label{euler2PN}\,.
\end{eqnarray}
Note that in this equation we have included also some 3PN contributions through the terms $p_{,\mu}/(\rho+p/c^2)$ ($\mu=t,i$)
appearing at 2PN. We did so because we know that the relativistic Euler 
equation is $a^\mu=- h^{\mu \nu}p_{,\nu}/(p+\rho c^2)$, so those terms are actually correct.

Combining the energy and rest-mass conservation equations, we obtain
\begin{multline}\label{Udot2PN}
\partial_t U=\ -(p+U){v^{j}_{ }}_{,j}-{U}_{,j}v^{j}
\\
+ \frac{p+U}{\rho c^2}\left({p}_{,j}v^{j}_{ }+4\rho{}{\phi{}}_{,j}v^{j}_{\  }+3\rho{}{\phi{}}_{,t}\right)
+{\cal O}\left(\frac{1}{c^4}\right)\,.
\end{multline}

Finally, using the equations derived in this paragraph, it is possible to show that
$\nabla^2 (\partial_i \omega^i)={\cal O}(1/c^4)$, $\nabla^2 (\partial_i \chi^{ij})={\cal O}(1/c^4)$ and $\nabla^2 \chi^{i}_i={\cal O}(1/c^4)$. Like in the 1PN case,
this implies that the gauge conditions $\partial_i \omega^i=\partial_i \chi^{ij}=\chi^i_i=0$ are automatically satisfied, at 2PN order, when imposed asymptotically far
away from the source.

\section{The equations at 2.5PN order: gravitational-wave radiation reaction}\label{sec2.5PN}

From the equations derived in the previous section, one can obtain the leading-order dissipative contribution
to the dynamics, i.e. the effect of gravitational-wave emission, which appears at 2.5PN order. More precisely, gravitational waves are encoded in the TT perturbation
$\chi_{ij}$, which at linear order (i.e. neglecting quadratic terms in the metric perturbations) is known to satisfy the wave 
equation~\cite{bardeen_pert,scott_lect}
\begin{equation}
\Box \chi_{ij} = -16 \pi \sigma_{ij}\,,
\end{equation}
where $\sigma_{ij}= P^k_i P^l_j T_{kl}-P_{ij} P^{kl} T_{kl}/2$ 
and $P_{ij}=\delta_{ij}-\nabla^{-2} \partial_i\partial_j$ is the transverse projector.
Clearly, the 2PN version of this equation [eq.~\eqref{eq:nabla2chi_ij}] does not contain the second time derivative $\partial_t^2 \chi_{ij}$ appearing
in the ``box'' operator $\Box=-\partial^2_t/c^2+\delta^{ij}\partial_i\partial_j$, because in the PN approximation time derivatives 
are suppressed by a factor $1/c$ relative to spatial derivatives. 
However, neglecting the time derivatives of quantities satisfying wave equations is subtle. For instance,
when $\chi_{ij}$ describes a gravitational wave with wavelength $\lambda$ propagating in the direction $\boldsymbol{n}$
at a distance $r\gg\lambda$ from the source (i.e. ``far'' from the source, in the so-called ``wave zone''),
 time derivatives are $\partial_t\chi_{ij}(\boldsymbol{x}-c t \boldsymbol{n})\approx - c\, n^k \partial_k\chi_{ij}$. The extra factor $c$ therefore
cancels the factor $1/c$ that accompanies time derivatives in the PN expansion, and
one cannot neglect time derivatives. 

Although we will show how to obtain an approximate solution for $\chi_{ij}$ in the wave zone at the end of this section [cf. eq.~\eqref{gw2}],
for the purposes of this work (which aims at evolving fluid configurations in the presence of black holes) 
it is actually more important to solve for $\chi_{ij}$ inside/near the source (i.e. in the
``near zone'' $r\ll\lambda$), because that is the regime that gives rise to the backreaction of gravitational waves on 
the source's dynamics. As we will now show, in the near zone the time derivatives (i.e. the retardation effects due to
the wave equation that $\chi_{ij}$ satisfies) will indeed cause the appearance of a 2.5PN dissipative radiation-reaction force.

Reinstating the time derivatives of $\chi_{ij}$, eq.~\eqref{eq:nabla2chi_ij} becomes~\footnote{In principle, 
the source $S_{ij}$ could contain terms depending on $\partial_t\chi_{ij}/c$. However, those terms are actually of higher PN order, as one can
check a posteriori from the decomposition of $\chi_{ij}$ in a 2PN term and a 2.5PN term [eq.~\eqref{decomposition}], which we will derive later in this section.}
\begin{eqnarray}
&&
\Box\chi_{ij}=\frac{1}{c^2} S_{ij}+{\cal O}\left(\frac{1}{c^4}\right)\label{wave_eq}
\\
&&
S_{ij}\equiv\delta^{ij}(8\pi p+{\phi{}}_{,k}{\phi{}}_{,k}+8\pi\rho{}v^2-2{\phi{}}_{,tt}) +4{\phi{}}_{,i}{\phi{}}_{,j} \nonumber\\
&&
+8\phi{}{\phi{}}_{,ij}+2{\delta\psi{}}_{,ij}
-{\omega{}^{i}_{ }}_{,jt}-{\omega{}^{j}_{ }}_{,it}-16\pi\rho{}v^{i}_{\  }v^{j}_{ } \label{Sij_fluid}
\,,
\end{eqnarray}
and recalling the retarded Green function of the $\Box$ operator, 
\begin{equation}
G(t,\boldsymbol{x})=-\frac{1}{4\pi |\boldsymbol{x}|} \delta(c\,t-|\boldsymbol{x}|),
\end{equation}
which satisfies $\Box G(t,\boldsymbol{x})=\delta(t)\delta^{(3)}(\boldsymbol{x})$,
one immediately gets
\begin{align}\label{full_sol_GW}
&\chi_{ij}(t,x^i)=\nonumber \\& \frac{1}{c^2}\int  G(t-t',\boldsymbol{x}-\boldsymbol{x}') S_{ij} (t',\boldsymbol{x}') {\rm d}^4 x'
+{\cal O}\left(\frac{1}{c^4}\right)\nonumber \\&=
-\frac{1}{c^2}\int  \frac{S_{ij} (t-|\boldsymbol{x}-\boldsymbol{x}'|/c,\boldsymbol{x}')}{4\pi |\boldsymbol{x}-\boldsymbol{x}'|}  {\rm d}^3 x'+{\cal O}\left(\frac{1}{c^4}\right)\,.
\end{align}
If the source $S_{ij}(t,\boldsymbol{x})$ had a compact support (i.e. if it vanished for $|\boldsymbol{x}|>\bar{r}$, where $\bar{r}$ is some finite radius), we
could expand in orders 
of $1/c$, and recalling that $\nabla^2 (1/ |\boldsymbol{x}|) =-4 \pi \delta^{(3)}(\boldsymbol{x})$, write
\begin{multline}
\chi_{ij}(t,x^i) =\\-\frac{1}{c^2}\int  \frac{S_{ij} (t,\boldsymbol{x}')}{4\pi |\boldsymbol{x}-\boldsymbol{x}'|}  {\rm d}^3 x'
+\frac{1}{c^3} \partial_t\int  \frac{ S_{ij} (t,\boldsymbol{x}')}{4\pi}  {\rm d}^3 x'+{\cal{O}}\left(\frac{1}{c^4}\right)\\=
\frac{1}{c^2}\nabla^{-2}S_{ij}+\frac{1}{c^3} \partial_t\int  \frac{ S_{ij} (t,\boldsymbol{x}')}{4\pi}  {\rm d}^3 x'+{\cal{O}}\left(\frac{1}{c^4}\right)\,.
\end{multline}
near or inside the support of $S_{ij}$ (i.e., loosely speaking, ``near the source'').
At first sight, it would seem that $S_{ij}$ does \textit{not} have a compact support, as it involves for instance the potential $\phi$, which decays as $1/r$
far from the source. However, because $\chi^{ij}$ is the TT part of the metric perturbation, we can take the TT projection of eq.~\eqref{wave_eq}, and obtain
\begin{eqnarray}
&&
\Box\chi_{ij}=\frac{1}{c^2} S^{\rm TT}_{ij}+{\cal O}\left(\frac{1}{c^4}\right)\label{wave_eq_projected}
\\
&&
S^{\rm TT}_{ij}\equiv \pi^{ijkl} S_{kl}  = -8 \pi \pi^{ijkl} (2 \rho v^k v^l+ \rho X_{,kl})\,,
\nonumber\\&&
\pi^{ijkl}\equiv P^{ik} P^{jl} -\frac12 P^{kl} P^{ij} \,,
\end{eqnarray}
where we have introduced the potential  
\begin{equation}
X(t,\boldsymbol{x})\equiv \int d^3 x' |\boldsymbol{x}-\boldsymbol{x'}| \rho(t,\boldsymbol{x}')\,,
\end{equation} 
which satisfies
$\nabla^2 X = - 2 \phi [1+{\cal O}(1/c^2)]$. We note that the only terms in $S_{ij}$ that are not trivially 
zeroed-out by the TT projector $\pi^{ijkl}$ are
$-16 \pi \rho v^i v^j$ and $4{\phi{}}_{,i}{\phi{}}_{,j}+8\phi{}{\phi{}}_{,ij}$. The latter terms can be rewritten,
 up to total derivatives (which are zeroed-out by the TT projection), as $4 \phi{}{\phi{}}_{,ij}$, which in turn
we can write as $4 \phi{}{\phi{}}_{,ij}= -2 X_{,ij} \nabla^2\phi  [1+{\cal O}(1/c^2)]+$ total derivatives.

Noting that the TT projection commutes with the $\Box$ operator, we can write 
\begin{multline}\label{chi25}
\chi^{ij}=\frac{1}{c^2}\Box^{-1} (S^{\rm TT}_{ij})+{\cal O}\left(\frac{1}{c^4}\right) =\\
-\frac{8\pi}{c^2} \pi^{ijkl} \Box^{-1}\zeta_{kl}+{\cal O}\left(\frac{1}{c^4}\right),
\end{multline}
where \begin{equation}\label{eq:zeta}\zeta_{ij}\equiv 2 \rho v^i v^j+ \rho X_{,ij}\end{equation} has now compact support. We can then expand
\begin{align}
\Box^{-1} \zeta_{ij} &=-\int  \frac{\zeta_{ij} (t-|\boldsymbol{x}-\boldsymbol{x}'|/c,\boldsymbol{x}')}{4\pi |\boldsymbol{x}-\boldsymbol{x}'|}  {\rm d}^3 x'\label{inv_box_zeta}
\\&=\nabla^{-2}\zeta_{ij}+\frac{1}{c} \partial_t\int  \frac{ \zeta_{ij} (t,\boldsymbol{x}')}{4\pi}  {\rm d}^3 x'+{\cal{O}}\left(\frac{1}{c^2}\right)\nonumber\\&\label{zeta_expansion}
\end{align}
Because the 2PN term $-{8\pi} \pi^{ijkl}  \nabla^{-2}\zeta_{kl}/c^2=-{8\pi} \nabla^{-2}( \pi^{ijkl} \zeta_{kl})/c^2$ 
appearing in $\chi^{ij}$ [c.f. eqs.~\eqref{chi25} and~\eqref{zeta_expansion}] 
is simply the solution to eq.~\eqref{eq:nabla2chi_ij}, we can write $\chi_{ij}$ as the sum
of the (conservative) 2PN contribution, and a dissipative 2.5PN contribution due to gravitational-wave emission:
\begin{align}\label{decomposition}
&\chi_{ij}(t,x^i) = \chi_{ij}^{\rm 2PN}(t,x^i)+\chi_{ij}^{\rm 2.5PN}(t,x^i)\\
&\chi_{ij}^{\rm 2.5PN}(t,x^i)=-\frac{8\pi}{c^3}  \pi^{ijkl}  \partial_t\int   \frac{\zeta_{kl} (t,\boldsymbol{x}')}{4\pi}  {\rm d}^3 x'\nonumber\\
&\quad=-\frac{8\pi}{c^3}  \partial_t\int   \frac{\bar{\zeta}^{ij}(t,\boldsymbol{x}') }{4\pi}  {\rm d}^3 x'\,,\label{gw1}\\
&\quad\bar{\zeta}^{ij}=\zeta_{ij}-\frac13\delta_{ij} \zeta^k_{k}\,.
\end{align}
This additional term enters in the Euler equation~\eqref{euler2PN}, where it gives rise to a dissipative ``radiation-reaction'' force. An obvious problem
with this is that $\chi^{ij}$ enters eq.~\eqref{euler2PN} through its derivatives (and in particular its time derivatives), and according to eq.~\eqref{gw1} $\chi^{ij}$ already depends on
the time derivatives of the velocity. One therefore needs to reduce the number of time derivatives using either the equations of motion
at Newtonian order~\cite{BDS_hydro} or reduction of order techniques at the code level (e.g.~\cite{Simon:1990ic,Flanagan:1996gw}).

To compute, if so desired, the gravitational waveforms produced by the system, one can instead approximate
eq.~\eqref{inv_box_zeta} far away from the source by replacing $|\boldsymbol{x}-\boldsymbol{x}'|$ with the
distance $r$ to the source and obtain 
\begin{align}\label{gw2}
&\chi_{ij}(t,x^i)=-\frac{8\pi}{c^2} \pi^{ijkl} \Box^{-1}\zeta_{kl}\nonumber \\&\approx
\frac{1}{c^2}\frac{2}{r}\left(p^{ik} p^{jl}-\frac12 p^{ij} p^{kl}\right)\int{\bar{\zeta}_{ij} (t-r/c,\boldsymbol{x}')}  {\rm d}^3 x'\nonumber\\&
+{\cal O}\left(\frac{1}{c^4}\right)\,,
\end{align}
with $p^{ij}=\delta^{ij}-n^i n^j$ ($n^i=x^i/r$ being the unit radial vector).

\section{The Kerr black hole - fluid system}\label{secKERRFLUID}
In order to describe a system comprised of a fluid and a rotating black hole, we \textit{define}
\begin{gather}\label{repl1}
\phi=\phi_{\rm fluid}+\phi_{\rm Kerr}\,,\\
\psi=\psi_{\rm fluid}+\psi_{\rm Kerr}\,,\\
\omega_i=\omega^{\rm fluid}_i+\omega^{\rm Kerr}_i\,,\\
\chi_{ij}=\chi^{\rm fluid}_{ij}+\chi^{\rm Kerr}_{ij}\,,\label{repl4}
\end{gather}
where we denote with the index ``fluid'' the part of the perturbations
that disappears in the limit in which $\rho$ and $p$ go to zero, while
the index ``Kerr'' denotes the part of the perturbations that makes up (in the Poisson gauge)
the Kerr metric, which describes an isolated rotating black hole. We stress that these equations
do \textit{not} amount to assuming any sort of superposition between an isolated Kerr black hole
and the fluid perturbations, as the ``fluid'' part will too depend
on the presence of the black hole (and in particular, on its mass and spin). 
For simplicity, it is convenient to exploit the freedom of choosing a particular reference frame, so as to 
be able to  describe the black hole with an (unchanging) Kerr metric.  With this choice, the interaction between the fluid and
the black hole will be accounted for by suitable terms sourcing the gravitational potentials. 
This reference frame can be identified by imposing appropriate boundary conditions on the
fluid potentials, i.e. by requiring that the fluid exerts neither a force nor a torque on the black hole, 
so that the black hole neither moves nor its spin precesses. Note that this does not mean that the fluid -- black hole interaction is
not accounted for. It simply amounts  -- at a Newtonian level -- to adopting a reference
frame comoving with the black hole and whose axes follow the precession of the black-hole spin. 

The motion of the black hole and the precession of its spin will be governed, at leading order in the spin, 
by the Mathisson-Papapetrou-Pirani 
equation~\cite{Math,Papa51, Papa51spin, CPapa51spin,Pirani,Tul1,Tul2,Dixon},
which one can write as~\cite{Da.al.08}
\begin{gather}
m \frac{D u_\mu}{D\tau} =\frac12\frac{\epsilon^{\alpha\beta\lambda\rho}}{\sqrt{-g}} S_\alpha u_\beta u_\nu R^\nu_{\phantom{\alpha}\mu\lambda\rho}
+{\cal O}(S)^2,\label{papapetrou1}
\\
\frac{D S_\mu}{D\tau} = {\cal O}(S)^2\,,\label{papapetrou2}
\end{gather}
where $u^\mu$ is the black-hole 4-velocity, $m$ its mass, $\tau$ its proper time, $R^\nu_{\phantom{\alpha}\mu\lambda\rho}$ and $g$ the Riemann tensor and metric determinant
of the ``external'' geometry (generated by the fluid) in which the black hole moves, and $S^\mu$ is the spin, which we assume to satisfy the ``covariant''
spin-supplementary condition $S_\mu u^\mu=0$. In order to determine the boundary conditions of the perturbations $\phi_{\rm fluid}$, $\psi_{\rm fluid}$,
$\omega^{\rm fluid}_i$ and $\chi^{\rm fluid}_{ij}$ at the black hole's position, such that the black hole does not move nor its spin precess, we can simply evaluate eqs.~\eqref{papapetrou1}
and~\eqref{papapetrou2} with the metric
\begin{multline}\label{FW}
ds^2=-\left(1+\frac{2}{c^2}a_i x^i\right) (d{x^0})^2+ \delta_{ij} dx^i dx^j \\+\frac{2}{c} \epsilon_{ikl} \Omega^k x^l dx^i dx^0 +{\cal O}(x)^2
\end{multline}
which locally describes a generic spacetime in Fermi-Walker local coordinates, i.e. in the reference frame of an observer (located at the origin) moving with
acceleration $a^i$ and whose axes precess with instantaneous angular velocity $\Omega^i$~\cite{FWref}. By replacing this metric in eqs.~\eqref{papapetrou1} and~\eqref{papapetrou2} 
and imposing that the
black hole does not move ($u^\mu=\delta^\mu_0$ and therefore $S_0=0$ from the spin supplementary condition) and its spin does not precess ($S_i$= const), 
we find that we can 
set the angular velocity to zero at linear order in the spin, i.e. $\Omega^{i}={\cal O}(S)^2$,
and
choose
\begin{equation}\label{BC}
m a_i=-\frac{c}{2}{\epsilon^{0jkl}} S_j R^0_{\phantom{\alpha}ikl}+{\cal O}(S)^2\,.
\end{equation}
One might then conclude that in order to achieve the desired conditions, 
we should impose $\partial_i\phi_{\rm fluid}=a_i$ (with $a_i$ given by eq.~\eqref{BC})
and $\partial_\mu\psi_{\rm fluid}=\psi_{\rm fluid}=\partial_\mu\omega^j_{\rm fluid}=\omega^j_{\rm fluid}=\partial_\mu\chi^{kl}_{\rm fluid}=
\chi^{kl}_{\rm fluid}=\partial_t\phi_{\rm fluid}=0$ at the black hole's position, so that the fluid-generated metric near the black hole matches eq.~\eqref{FW}.

These boundary conditions would be, however, overly restrictive. 
For instance, one can rewrite eq.~\eqref{FW} in a different
 coordinate system, by rescaling the time and by rotating and rescaling the spatial axes. When one does so, it is
easy to get convinced that a more generic set of boundary conditions under which the black hole does not move nor its spin precesses
 is given by $\partial_i\phi_{\rm fluid}=a_i (1+2 \phi_{\rm fluid}/c^2)$ and
$\partial_\mu\psi_{\rm fluid}=\partial_\mu\omega^j_{\rm fluid}=\partial_\mu\chi^{kl}_{\rm fluid}=\partial_t\phi_{\rm fluid}=0$ at the black hole's position,
with no conditions on the values of the fluid-generated metric perturbations at the black hole, except that they be constant in time (because all the time
derivarives of the metric perturbations at the black hole's location vanish).

An even more general approach is to insert the metric \eqref{metric2} into eqs.~\eqref{papapetrou1} and \eqref{papapetrou2},
and require that $S_i$ and $u^i=0$ remain constant.\footnote{Note that requiring that the coordinate components of the spin $S_i$ remain constant
is actually a stronger constraint than simply requiring no spin precession. For instance, the coordinate
components of the spin may vary simply due to a change of the conformal
factor $1-2\psi_{\rm fluid}/c^2$ [cf. eq.~\eqref{metric2}], even in the absence of precession. However, if we allowed such situations,
we would need to rescale the coordinates of the Kerr metric in eqs.~\eqref{repl1} -- \eqref{repl4} at each timestep, which would be rather impractical.}   
Doing so, one obtains the conditions
\begin{gather}
m c \Gamma^i_{00} =-\frac12 \sqrt{-g}\, {\epsilon_{0\alpha\lambda\rho}} S^\alpha R_{0}^{\phantom{\alpha}i\lambda\rho}\,,\\
\Gamma^i_{j 0} S_i=0\,.
\end{gather}
We note, however, that in many physically relevant situations the fluid configuration is typically significantly
less massive than the black hole.
As a result, the fluid barely influences the black-hole position and the orientation of its spin, i.e. the boundary conditions given above
will be approximately satisfied anyway.

The Kerr part of the perturbations in the Poisson gauge may be read off 
the Kerr metric in ADM-TT coordinates~\cite{Hergt:2007ha},
\begin{equation}\label{downmetric}
g_{\mu\nu}=
\begin{pmatrix}
-\alpha^2+\beta_{i}\beta^{i}  & -\beta_{i}\\
-\beta_{j} & \gamma_{ij}
\end{pmatrix}\,,
\end{equation}
where $\gamma^{ik}\,\gamma_{kj}=\delta^i_j$ and $\beta^i=\gamma^{ik}\, \beta_k$.
We define ${n}^i\equiv {x}^i/r$, denote  the mass of the Kerr black hole by $M$ and its spin by $\boldsymbol{S}_{\rm Kerr}$, 
and introduce a dimensionless three-vector 
\begin{equation}\label{spin_par}
\boldsymbol{\chi}\equiv \frac{c \boldsymbol{S}_{\rm Kerr}}{M^2}\,,
\end{equation}
whose norm $\chi=|\boldsymbol{\chi}|\leq1$ represents the spin parameter of the black hole.
The lapse function is given by~\cite{Hergt:2007ha}
\begin{eqnarray}
\alpha &=& 1-\frac{M}{rc^2}+\frac{1}{2}\frac{M^2}{r^2c^4}-\frac{1}{4}\frac{M^3}{r^3c^6} \nonumber \\
&& + \frac{1}{8}\frac{M^4}{r^4c^8} + \frac{1}{2}\frac{M^3[3(\boldsymbol{\chi}\!\cdot\!\boldsymbol{n})^2-\chi^2]}{r^3 c^6} \nonumber \\
&& +\frac{1}{2}\frac{M^4[5\chi^2-9(\boldsymbol{\chi}\!\cdot\!\boldsymbol{n})^2]}{r^4c^8}+{\cal O}\left(\frac{1}{c^{10}}\right),\label{laadm}
\end{eqnarray}
the shift vector is given by
\begin{multline}
\beta^{i}= \Bigg\{\frac{2M^2}{r^2 c^4}-\frac{6M^3}{r^3 c^6}
+\frac{21}{2}\frac{M^4}{r^4 c^8} \\-\frac{M^4 [5(\boldsymbol{\chi}\!\cdot\!\boldsymbol{n})^2
-\chi^2]}{r^4 c^8}\Bigg\}\epsilon^{ijk}\chi_{j}n_{k}+{\cal O}\left(\frac{1}{c^{10}}\right),\label{shadm}
\end{multline}
(where $\epsilon_{ijk}=\epsilon^{ijk}$ is the Levi-Civita symbol, with
$\epsilon_{123}=\epsilon^{123}=1$),
and the spatial metric $\gamma_{ij}$ is given by
\begin{equation}\label{gammametric}
\gamma_{ij}= {A}\delta_{ij}+ h_{ij}^{\rm TT}+{\cal O}\left(\frac{1}{c^{10}}\right)\,,
\end{equation}
where the quantities $A$ and $h_{ij}^{\rm TT}$ are defined as
\begin{eqnarray}
A &=& \left(1+\frac{M}{2r c^2}\right)^4+\frac{M^3[\chi^2-3(\boldsymbol{\chi}\!\cdot\!\boldsymbol{n})^2]}{r^3 c^6}  \nonumber\\
&& +\frac{1}{2}\frac{M^4\chi^2}{r^4c^8}-\frac{3M^4(\boldsymbol{\chi}\!\cdot\!\boldsymbol{n})^2}{r^4 c^8}\,,
\end{eqnarray}
\begin{eqnarray}
 h_{ij}^{\rm TT} &=& -\frac{7}{2}\frac{M^4\chi^2}{r^4c^8}\delta_{ij}+7\frac{M^4(\boldsymbol{\chi}\!\cdot\!\boldsymbol{n})^2}{r^4c^8}\delta_{ij} \nonumber \\
&& + 7\frac{M^4\chi^2n_{i}n_{j}}{r^4c^8}-21\frac{M^4(\boldsymbol{\chi}\!\cdot\!\boldsymbol{n})^2n_{i}n_{j}}{r^4c^8} \nonumber\\
&& +\frac{7}{2}\frac{M^4\chi_{i}\chi_{j}}{r^4 c^8}\,.\label{hij}
\end{eqnarray}
It is easy to check that $\partial_i g_{0i}=\partial_j h_{ij}^{\rm TT}=\delta^{ij}h_{ij}^{\rm TT}={\cal O}(1/c^{10})$, which are
exactly the Poisson gauge conditions. From these equations one thus gets
\begin{align}
&\phi_{\rm Kerr}= \frac{c^2}{2}(\alpha^2-\beta_{i}\beta^{i}-1) = \\ &\quad
-\frac{M}{r} + \frac{1}{c^2}\frac{M^2}{r^2} - \frac{M^3}{2 c^4} \frac{[\chi^2-3(\boldsymbol{\chi}\!\cdot\!\boldsymbol{n})^2]}{r^3}
\\ &\quad- \frac34 \frac{M^3}{r^3 c^4}+{\cal O} \left(\frac{1}{c^6}\right)\, ,\nonumber\\ &
\delta\psi_{\rm Kerr} = \frac{c^4}{2} (1-A)-c^2\phi_{\rm Kerr}  =-\frac{7M^2}{4r^2}+{\cal O} \left(\frac{1}{c^2}\right)\,,\\ &
\omega^i_{\rm Kerr}=-c^3 \beta_i\equiv\frac{1}{c}\widetilde{\omega}^i_{\rm Kerr}\,,\\ &
\widetilde{\omega}^i_{\rm Kerr}=\left(\frac{2 M^2}{r^2}-\frac{2 M^3}{r^3 c^2}\right)\epsilon_{ijk} n^j \chi^k+{\cal O} \left(\frac{1}{c^4}\right)\,,\\ &
\chi^{ij}_{\rm Kerr}= c^2 h_{ij}^{\rm TT} = {\cal O} \left(\frac{1}{c^6}\right)\,.\\ &
\end{align}
where we have rescaled $\omega^i_{\rm Kerr}$ by a factor $1/c$ to $\widetilde{\omega}^i_{\rm Kerr}$ to highlight the 
fact that $\omega^i_{\rm Kerr}$ appears at 1.5PN order for spinning black holes or compact objects.
Using the fact that the Kerr metric is a solution to the vacuum Einstein equation, at 1.5PN the generalized Poisson equation becomes
\begin{multline}\label{nabla2phi1.5PN}
\nabla^2 \phi_{\rm fluid}=4\pi \left(3 \frac{p}{c^2} +\rho\right)+\frac{2}{c^2}{\phi_{\rm fluid}}_{,i}{\phi_{\rm fluid}}_{,i}\\
+\frac{4}{c^2}{\phi_{\rm fluid}}_{,i}{\phi_{\rm Kerr}}_{,i}
+8 \pi\rho \left(\frac{v}{c}\right)^2-\frac{3}{c^2}{\phi_{\rm fluid}}_{,tt}+{\cal O}\left(\frac{1}{c^4}\right)\,,
\end{multline}
the equation for the gravitomagnetic potential becomes
\begin{equation}\label{nabla2omega1.5PN}
\nabla^2 \omega_{\rm fluid}^i= 4 (4 \pi\rho v^i+{\phi_{\rm fluid}}_{,ti})+{\cal O}\left(\frac{1}{c^2}\right)\,,
\end{equation}
while the 1.5 PN energy-conservation, Euler and rest-mass conservation equations are given by 
Eqs.~\eqref{rhodot1PN}, \eqref{vidot1PN} and \eqref{Udot1PN}, simply by replacing the metric
perturbations with Eqs.~\eqref{repl1}--\eqref{repl4}:
\begin{multline}
{\rho}_{,t}+{\rho}_{,i}v^{i}+\left(\frac{p}{c^2}+\rho\right){v^{i}}_{,i}-\frac{p_{,i} v^i}{c^2} 
\\- \frac{\rho}{c^2} [4 (\phi_{{\rm fluid},i}+\phi_{{\rm Kerr},i}) v^i+3 \phi_{{\rm fluid},t}]={\cal O}\left(\frac{1}{c^4}\right)\,,
\end{multline}
\begin{equation}
\partial_t U=\ -\left(p+U\right){v^{j}_{ }}_{,j}-{U}_{,j}v^{j}+{\cal O}\left(\frac{1}{c^2}\right)\,.
\end{equation}
and
\begin{multline}\label{euler1.5PN}
{v^{i}}_{,t}+{v^{i}}_{,a}v^{a} =
-{\phi_{\rm fluid}}_{,i}-{\phi_{\rm Kerr}}_{,i}
\\
-\frac{{p}_{,i}}{\rho+p/c^2}  
-\frac{4}{\rho c^2+p}(\phi_{\rm fluid}+\phi_{\rm Kerr}){p}_{,i}
\\-\frac{2}{c^2} (\phi_{{\rm fluid}} +\phi_{{\rm Kerr}}) (\phi_{{\rm fluid},i}+ \phi_{{\rm Kerr},i})\\
 - \left(\frac{v}{c}\right)^2 \left(-\frac{p_{,i}}{\rho+p/c^2}+\phi_{{\rm fluid},i}+ \phi_{{\rm Kerr},i}\right)\\
+\frac{v^i}{c^2}  \left[4(\phi_{{\rm fluid},j}+\phi_{{\rm Kerr},j}) v^j+3\phi_{{\rm fluid},t}-\frac{p_{,t}}{\rho+p/c^2}\right]\\
+\frac{1}{c^2}\left(-{{\omega^{i}_{\rm fluid}}}_{,t}
-{{\omega^{i}_{\rm fluid}}}_{,a}v^{a}
+{{\omega^{a}_{\rm fluid}}}_{,i}v^{a}\right)\\
+\frac{1}{c^3}\left(
-{{\widetilde{\omega}}^{i}_{{\rm Kerr},a}}v^{a}
+{\widetilde{\omega}^{a}_{{\rm Kerr},i}}v^{a}
\right)+{\cal O}\left(\frac{1}{c^4}\right)
\,.
\end{multline}
Again, in the Euler equation we have replaced the expressions $p_{,\mu}/\rho$ ($\mu=t,i$)
appearing at 1PN with $p_{,\mu}/(\rho+p/c^2)$. This
introduces corrections at 2 PN, but we know that the relativistic Euler 
equation is $a^\mu=- h^{\mu \nu}p_{,\nu}/(p+\rho c^2)$, so those terms are actually correct,
as can be seen explicitly from eq.~\eqref{vidot2PN}.

Similarly, at 2.5PN order the equations for the metric perturbations produced by the fluid become
\begin{multline}\label{nabla2phi2.5}
\nabla^2 \phi_{\rm fluid}=4\pi \left(3 \frac{p}{c^2} +\rho\right)+\frac{2}{c^2}{\phi_{\rm fluid}}_{,i}{\phi_{\rm fluid}}_{,i}\\
+\frac{4}{c^2}{\phi_{\rm fluid}}_{,i}{\phi_{\rm Kerr}}_{,i}
+8 \pi\rho \left(\frac{v}{c}\right)^2-\frac{3}{c^2}{\phi_{\rm fluid}}_{,tt}\\
+\frac{1}{c^4}\Big[-16\pi\rho(\phi_{\rm fluid}+\phi_{\rm Kerr})^{2}-8\pi\rho\,(\delta\psi_{\rm fluid}+\delta\psi_{\rm Kerr})\\
+{\phi_{\rm fluid}}_{,i}{\delta\psi_{\rm Kerr}}_{,i}+{\phi_{\rm Kerr}}_{,i}{\delta\psi_{\rm fluid}}_{,i}+{\phi_{\rm fluid}}_{,i}{\delta\psi_{\rm fluid}}_{,i}\\
+{\phi_{\rm fluid}}_{,i}{\omega^{i}_{\rm fluid}}_{,t}+{\phi_{\rm Kerr}}_{,i}{\omega^{i}_{\rm fluid}}_{,t}
-\frac12 {\omega_{\rm fluid}^{i}}_{,j}{\omega_{\rm fluid}^{i}}_{,j}\\
+\frac12 {\omega_{\rm fluid}^{i}}_{,j}{\omega_{\rm fluid}^{j}}_{,i}
+8\pi[p-4\rho{}(\phi_{\rm fluid}+\phi_{\rm Kerr})]v^2+8\pi\rho v^4\\
+{\phi_{\rm fluid}}_{,ij}\chi_{\rm fluid}^{ij}+{\phi_{\rm Kerr}}_{,ij}\chi_{\rm fluid}^{ij}
-3{\delta\psi_{\rm fluid}}_{,tt}\Big]\\
+\frac{1}{c^5}\Big(
- {\widetilde{\omega}_{{\rm Kerr},j}^{i}}{\omega_{\rm fluid}^{i}}_{,j}
+ {\widetilde{\omega}_{{\rm Kerr},j}^{i}}{\omega_{\rm fluid}^{j}}_{,i}\Big)
+{\cal O}\left(\frac{1}{c^6}\right)\,,
\end{multline}
\begin{multline}\label{nabla2psi2.5}
\nabla^2\delta\psi_{\rm fluid}=-12p\pi-16\pi\rho{}(\phi_{\rm fluid}+
\phi_{\rm Kerr})-{7\over 2}{\phi_{\rm fluid}}_{,j}{\phi_{\rm fluid}}_{,j}\\
-7{\phi_{\rm fluid}}_{,j}{\phi_{\rm Kerr}}_{,j}-4\pi\rho{}v^2
+3{\phi_{\rm fluid}}_{,tt}+{\cal O}\left(\frac{1}{c^2}\right)\,,
\end{multline}
\begin{align}\label{nabla2chi2.5}
&\nabla^2\chi^{\rm fluid}_{ij}=\frac{1}{c^2}S_{ij}+{\cal O}\left(\frac{1}{c^4}\right)\,,\\\nonumber
&S_{ij}\equiv\delta^{ij}\big(8\pi p+{\phi_{\rm fluid}}_{,k}{\phi_{\rm fluid}}_{,k}
+2{\phi_{\rm fluid}}_{,k}{\phi_{\rm Kerr}}_{,k}\\\nonumber&+8\pi\rho{}v^2-2{\phi_{\rm fluid}}_{,tt}\big)
+4{\phi_{\rm fluid}}_{,i}{\phi_{\rm fluid}}_{,j}+8{\phi_{\rm fluid}}_{,i}{\phi_{\rm Kerr}}_{,j}\\&\nonumber
+8\phi_{\rm fluid}{\phi_{\rm fluid}}_{,ij}+8\phi_{\rm Kerr}{\phi_{\rm fluid}}_{,ij}+8\phi_{\rm fluid}{\phi_{\rm Kerr}}_{,ij}\\&
+2{\delta\psi_{\rm fluid}}_{,ij}
-{\omega{}^{i}_{\rm fluid }}_{,jt}-{\omega{}^{j}_{\rm fluid }}_{,it}-16\pi\rho{}v^{i}_{\  }v^{j}\,,\label{Sij_kerr}
\end{align}
and
\begin{multline}\label{nabla2omega2.5}
\nabla^2 \omega_{\rm fluid}^i= 4 (4 \pi\rho v^i+{\phi_{\rm fluid}}_{,ti})
+\frac{2}{c^2}\Big[{\phi_{\rm fluid}}_{,j}{\omega{}^{j}_{\rm fluid}}_{,i}\\+{\phi_{\rm Kerr}}_{,j}{\omega{}^{j}_{\rm fluid}}_{,i}
-{\phi_{\rm fluid}}_{,ij}\omega{}^{j}_{\rm fluid}-{\phi_{\rm Kerr}}_{,ij}\omega{}^{j}_{\rm fluid}
\\
+2\Big({\phi_{\rm fluid}}_{,t}{\phi_{\rm fluid}}_{,i}+{\phi_{\rm fluid}}_{,t}{\phi_{\rm Kerr}}_{,i}
+{\delta\psi_{\rm fluid}}_{,it}
+4p\pi v^{i}\\-16\pi\rho(\phi_{\rm fluid}+\phi_{\rm Kerr})v^{i}
+4\pi\rho{}v^2 v^{i}_{ }+2\pi\rho\omega_{\rm fluid}^{i}\Big)\Big]
 \\+\frac{2}{c^3}\Big(
 {\phi_{\rm fluid}}_{,j}{\widetilde{\omega}^{j}_{{\rm Kerr},i}}-{\phi_{\rm fluid}}_{,ij}\widetilde{\omega}^{j}_{\rm Kerr}
 +4\pi\rho\widetilde{\omega}_{\rm Kerr}^{i}\Big)
 +{\cal O}\left(\frac{1}{c^4}\right)\,,
\end{multline}
The dissipative part of the metric perturbations is described by suitable extensions to 
eqs.~\eqref{gw1} and \eqref{gw2}. The presence of the black hole introduces a modification to
the source $\zeta_{ij}$ [eq.~\eqref{eq:zeta}], which picks up terms
due to the black hole-fluid interaction. More specifically, applying the same
procedure of Section~\ref{sec2.5PN} to eq.~\eqref{nabla2chi2.5}, we find 
\begin{equation}\label{eq:zeta_new}
\zeta_{ij}= 2 \rho v^i v^j+ \rho X_{,ij}+2 M X_{,ij} \delta^{(3)}(\boldsymbol{x})
\end{equation}
(where, as usual, $n^i=x^i/r$). Note that the extra term
is nothing but the ``mass density'' of the black hole, $\rho_{\rm Kerr}=M \delta^{(3)}(\boldsymbol{x})$,
multiplied by $2 X_{,ij}$. This term is the only additional surviving contribution arising
when applying the procedure of Section~\ref{sec2.5PN} to the terms in expression (\ref{Sij_kerr}) that do not appear
already in equation (\ref{Sij_fluid}). Similarly, gravitational waveforms may be obtained via eq.~\eqref{gw2}, with
the source $\zeta_{ij}$ again given by eq.~\eqref{eq:zeta_new}. 

Finally, as in the 1.5PN case outlined above, 
the 2.5PN energy-conservation equation, Euler equation and rest-mass conservation equation are 
given by Eqs.~\eqref{rhodot2PN}, \eqref{vidot2PN} and \eqref{Udot2PN}, simply by replacing the metric
perturbations with Eqs.~\eqref{repl1}--\eqref{repl4}. 

\section{Possible models for numerical implementation}\label{secMODELS}

Based on the results of the previous section, there are several possible implementations of
a PN scheme to describe the fluid -- black hole system. As far as the conservative dynamics is concerned 
we can either 
 truncate the PN series at 1.5PN and therefore solve only the generalized Poisson equation 
 \eqref{nabla2phi1.5PN} for $\phi_{\rm fluid}$ and eq.~\eqref{nabla2omega1.5PN} for the 
``frame-dragging'' potential 
$\omega_i^{\rm fluid}$, or truncate the series at 2.5PN and therefore solve
 eqs~\eqref{nabla2phi2.5}, \eqref{nabla2psi2.5}, \eqref{nabla2chi2.5} and \eqref{nabla2omega2.5}
 for $\phi_{\rm fluid}$, $\delta\psi_{\rm fluid}$, $\chi_{ij}^{\rm fluid}$ and $\omega_i^{\rm fluid}$.

As far as the dissipative dynamics and gravitational waves are concerned, one can use 
respectively eq.~\eqref{gw1} 
and eq.~\eqref{gw2}, with
the source $\zeta_{ij}$ given by eq.~\eqref{eq:zeta_new}. 

Finally, for the fluid's equations of motion (the Euler, energy-conservation, and rest-mass conservation equations), we can either use the Taylor expanded forms presented in the 
previous sections (truncated at 1PN/2PN order, in the absence of a black hole, or at 1.5PN/2.5PN order if a spinning black hole is present), or the full unexpanded forms
\begin{gather}
u^\nu \nabla_\nu u^\mu =-\frac{(g^{\mu\nu}+u^\mu u^\nu)\partial_{\nu} p}{p+\rho c^2}\,\\
u^\mu \partial_\mu \rho=-\left(\frac{p}{c^2}+\rho\right) \nabla_\mu u^\mu\label{resummed_euler}\,\\
\nabla_{\alpha} (\mu u^\alpha)=0\,.
\end{gather}

As a result we have four possible models: 1.5PN conservative + 2.5PN dissipative + Taylor expanded  equations of motion (``1.5PNc+2.5PNd+Taylor EOM''); 
   1.5PN conservative + 2.5PN dissipative + unexpanded  equations of motion (``1.5PNc+2.5PNd+resummed EOM'');   
   2.5PN conservative + 2.5PN dissipative +  Taylor expanded  equations of motion (``2.5PNc+2.5PNd+Taylor EOM''); 
     2.5PN conservative + 2.5PN dissipative + unexpanded  equations of motion (``2.5PNc+2.5PNd+resummed EOM'').
In order to gauge the faithfulness of each of these models relative to the exact general-relativistic result, we 
look at isolated (i.e. spherically symmetric and static) stars and isolated spinning black-holes. (Clearly, neither of these tests depends on the 
dissipative PN dynamics, but we stress that our choice of truncating the dissipative dynamics at leading order is akin to 
using the quadrupole formula in place of calculating the exact gravitational-wave fluxes. This is a standard choice 
in approaches that do not implement full General Relativity [see e.g. Refs.~\cite{r_instability,2009ApJ...705..844G,2010AA...514A..66R,2006Sci...312..719P}]).

\begin{figure}
\includegraphics[width=6cm,angle=-90]{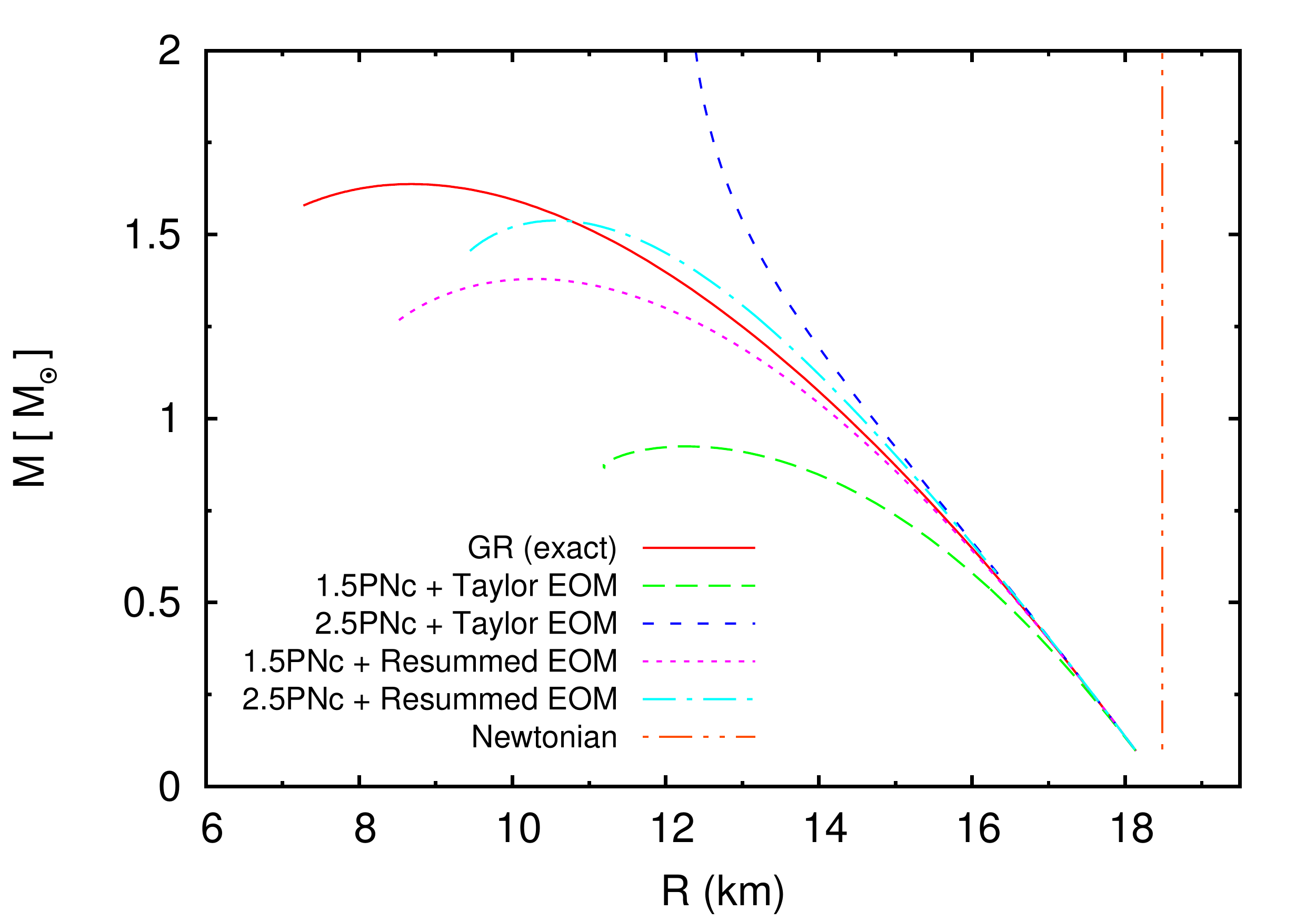}
\caption{\footnotesize The gravitational mass vs radius (in isotropic coordinates) for static and 
spherically symmetric stars (described by a perfect fluid obeying a polytropic equation of state 
with $K=100 G^3 M_\odot^2/c^6$ and $\Gamma=2$), 
for the various models presented in Sec.~\ref{secMODELS}, for Newtonian theory, and for General 
Relativity.\label{figure_mass}}
\end{figure}

\begin{figure}
\includegraphics[width=6cm,angle=-90]{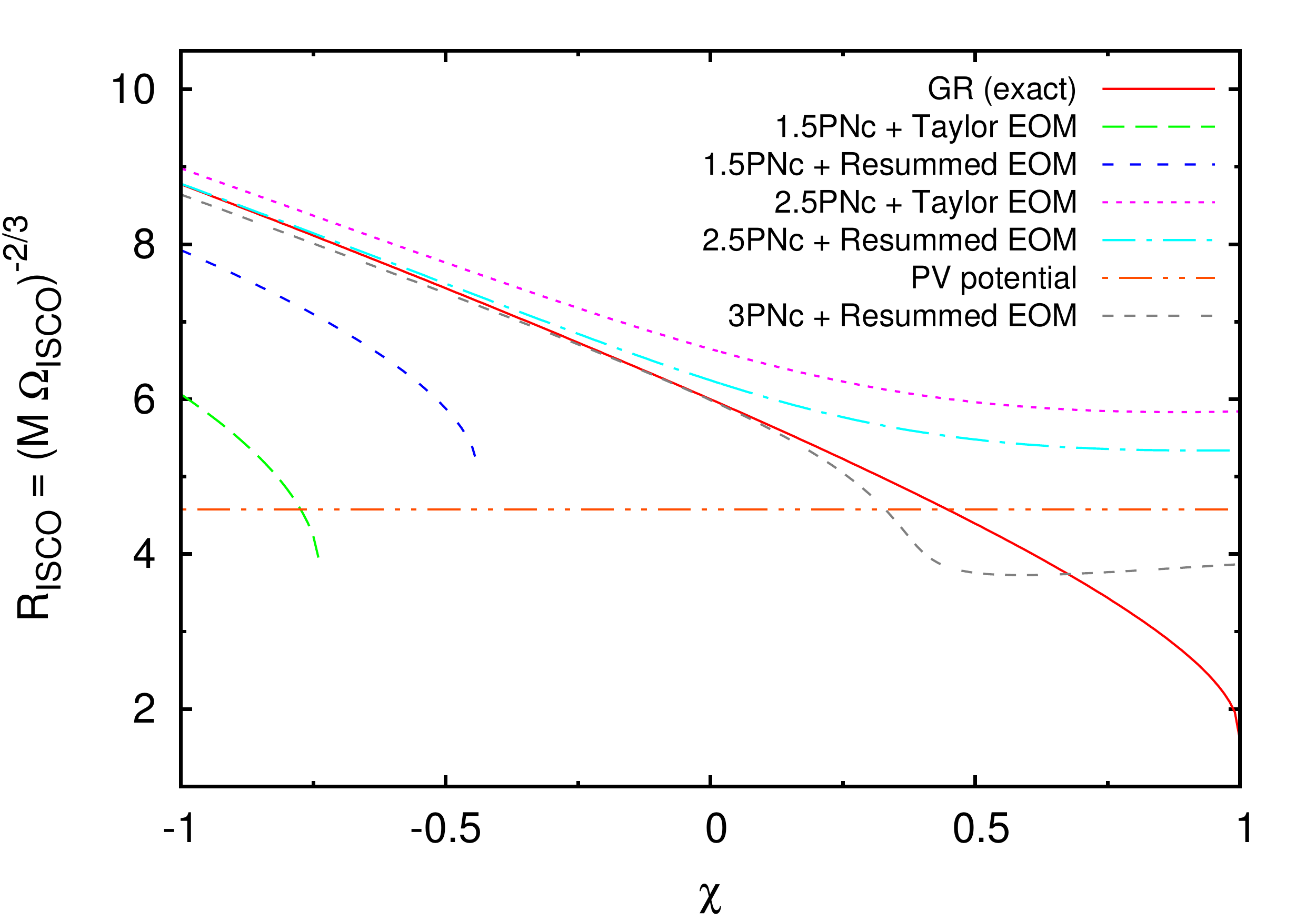}
\caption{\footnotesize The gauge-invariant ISCO radius $R_{_{\rm ISCO}}=(M\Omega_{_{ISCO}})^{-2/3}$ of an 
isolated rotating black hole, 
for the various models presented in Sec.~\ref{secMODELS}, for the pseudo-Newtonian PV potential, 
and for General Relativity.\label{figure_isco}}
\end{figure}

We model isolated stars with  a perfect fluid satisfying a polytropic equation of state
with $K=100 G^3 M_\odot^2/c^6$ and $\Gamma=2$ (which provides a good approximation for cold neutron stars).
Figure~\ref{figure_mass} shows the gravitational mass $M$ of the stars vs their radius $R$ in
isotropic coordinates~\footnote{In the static, spherically symmetric case that
we consider here, $\chi_{\rm fluid}^{ij}=\omega_{\rm fluid}^{i}=0$ without loss of generality, so one only has to solve
for $\phi_{\rm fluid}$ and $\delta\psi_{\rm fluid}$, and the spatial metric is conformally flat.} for each of the possible models together with the result obtained
in full General Relativity.  As can be seen the Newtonian prediction for this particular polytropic 
index is $R =$ const $\approx18.5$ km independent of the mass. This is clearly very far from the exact general-relativistic result, 
which is better approximated by our PN  schemes, and in particular by the ``1.5PNc+resummed EOM'' 
and ``2.5PNc+resummed EOM'' models.

In the case of black holes, we focus on the ISCO as calculated in each
model. The existence of an ISCO has profound astrophysical
implications. For instance, the ISCO regulates the radiative efficiency and the location of the inner edge of 
thin accretion disks, while in compact-object binaries, it marks the boundary where a quasi-adiabatic inspiral transitions into a plunge and merger phase.
We thus calculate the gauge-invariant ISCO 
radius $R_{_{\rm ISCO}}=(M\Omega_{_{ISCO}})^{-2/3}$, defined in terms of the ISCO orbital frequency $\Omega_{_{\rm ISCO}}$.
This measure of the ISCO radius is to be preferred to coordinate radii because it does not depend on
the coordinate system and is thus free of ambiguities (for instance, while the coordinate location of the ISCO in the PV potential
is correct in a particular coordinate system, the associated ISCO frequency is incorrect). 
To calculate the ISCO frequency we take the Kerr metric in ADM coordinates, 
as given in sec.~\ref{secKERRFLUID}, truncate it
at the appropriate (1.5PN or 2.5PN) order, and utilize it in the Euler equations (with the pressure set to zero), 
either Taylor expanded at the appropriate (1.5PN or 2.5PN) order, or in their ``resummed''
form~\eqref{resummed_euler}. (Of course, the Euler equation reduces to the geodesic equation when $p=0$). 
The ISCO location is then obtained by studying the stability of circular orbits under radial perturbations.
Because circular orbits are assumed to have $\dot\phi>0$ (i.e. to move in the positive $\phi$-direction), 
negative values of the spin-parameter projection $\chi$ denote configurations in 
which the orbit's angular momentum and the black-hole spin are anti-aligned.

As can be seen, the ``1.5PNc+Taylor EOM'' and ``1.5PNc+resummed EOM'' models do not perform well, because they are far from the general-relativistic result, and for $\chi\gtrsim-0.5$ they do not
seem to  present an ISCO at all. Better results are achieved by the ``2.5PNc+Taylor EOM'' and ``2.5PNc+resummed EOM'' models, which 
present an ISCO over the whole spin range $\chi\in [-1,1]$. We stress that spin-effects (``frame-dragging'')
are completely absent for the PV potential, whose expression $V(r)=-M/(r-2 M/c^2)$ has no spin dependence.
Nevertheless, it is rather clear that none of our PN schemes can reproduce the exact general-relativistic result at high spins $\chi\sim1$. While this 
behavior is clearly unsatisfactory, it is common to essentially all
approximation schemes based on PN theory (even if the PN dynamics is resummed into an EOB model, cf. e.g. Ref.~\cite{taracchini}). This is because PN theory necessarily fails when
$\chi\to1$, since in the extremal limit the Kerr ISCO coincides with a null generator of the horizon~\cite{ted_isco}, 
where the PN expansion breaks down as $v\sim c$, i.e. all terms in the PN expansion would be needed to obtain an accurate 
determination of the ISCO location. 
In our case, we recall that we are forced by our initial choice of using the Poisson gauge
to use the Kerr metric in ADM coordinates, which is only known through 3PN order~\cite{Hergt:2007ha}. 
(As mentioned earlier, the choice of the Poisson gauge is dictated by the need to minimize
the number of propagating degrees of freedom satisfying hyperbolic equations.)
The explicit form of
the 3PN Kerr metric in ADM coordinates is given by eqs.~\eqref{downmetric} -- \eqref{hij}, and we have attempted to use that form in the unexpanded (``resummed'') geodesic equation $a^\mu=0$ to
calculate the ISCO. The result is shown in Fig.~\ref{figure_isco} as ``3PNc+resummed EOM''. As can be seen, at high spins that curve still falls short of reproducing the exact Kerr ISCO, 
confirming that a precise determination of this quantity would require many more PN orders than currently available. 

\section{Conclusions}
In this work we have presented a PN formulation of the equations of motion for a black hole interacting with
a fluid configuration (e.g.~a star). Our approach
can in principle be implemented in existing Newtonian codes to
account for currently missing effects appearing at different PN orders (e.g. frame dragging,
black-hole spins, radiation reaction, etc.). Alternatively, it provides for a way to 
estimate the errors intrinsic to Newtonian approaches because of their neglecting of
PN effects. 
As illustrated in Fig. 1 and 2, our approach 
is approximate, 
but its performance improves as higher PN orders are considered (especially for black-hole spin
parameters $cJ/(GM^2)\lesssim0.5$). 
Future work will concentrate on the application of this 
approach in relevant systems.\\

\begin{acknowledgments}
We would like to thank Luc Blanchet, Thomas Janka, Eric Poisson, Oscar Reula, Stephan Rosswog, 
Enrico Ramirez-Ruiz, Olivier Sarbach, Eliot Quataert, and especially Guillaume Faye 
for helpful discussions.
E. B. acknowledges support from a CITA National Fellowship while at
the University of Guelph, and from the European Union's Seventh Framework Programme (FP7/PEOPLE-2011-CIG)
through the Marie Curie Career Integration Grant GALFORMBHS PCIG11-GA-2012-321608
while at the Institut d'Astrophysique de Paris.
We also acknowledge hospitality 
from the Kavli Institute for Theoretical Physics at UCSB (E.B. and L.L.)  and Perimeter
Institute (E.B.), where part of this work was carried out. This work was supported in part by the 
National Science Foundation under grant No. NSF PHY11-25915 (to UCSB) and NSERC through a Discovery Grant (to LL). 
Research at Perimeter Institute is supported through Industry Canada and by the Province of Ontario
through the Ministry of Research \& Innovation.
\end{acknowledgments}

\end{document}